**D.S. Belyakov, E.O. Kalinin, A.A. Konev, A.A. Shelupanov , A.K. Novokhrestov**

# Life cycle models and security threats to a microchip during its development and operation

The growth of Internet of Things devices has shown the need to develop the direction of information security in the field of development and operation of microcircuits, since modern information systems are built around the latter. This article presents the life cycle of secure chips used as a root of trust ( Root of Trust ) information systems. The main stages of the life cycle of protected microcircuits are described, namely, the life cycle models during development and during operation by the end user.
**Keywords:** safety , safety module , life cycle , protected microcontrollers.

In the context of the rapid growth of cyber threats , it is necessary to ensure the possibility of secure information exchange in automated systems. This is especially true for critical information infrastructures (CII) in backbone industries (for example, energy, rail and air transport, chemical production, banking systems, healthcare, nuclear and defense industries), where a successful cyber attack can have the most severe consequences [1- 2]. For this reason, in these industries, the creation of trusted automated systems is coming to the fore.

The key approach in creating trusted systems is the use of secure microcontrollers with a fully controlled life cycle [3-4] from the development of a crystal to the creation of devices based on them.

A secure microcontroller is a semiconductor device that, in addition to the processor core, includes additional hardware blocks to reduce the execution time of cryptographic operations (cryptographic accelerators) and implements protection measures that counteract threats aimed at confidential data through impact on the microcircuit.

In modern realities, microcircuits are used in Internet of things devices, where they are responsible for ensuring the security of data transmission channels [5] through the use of cryptographic mechanisms built into the microcircuit.

During its lifetime, a microcircuit can be in two states - in the state of development or operation. During development, the chip creation process and accompanying software (such as a development kit) must be protected. During operation, it is necessary to ensure the security of the process of using the microcircuit by both the developer and the end user of the device, which includes this microcircuit. The separation of the presented processes allows us to consider each individual threat in more detail, as well as to formalize the life cycle models [6-7].

A similar approach to the formalization of the life cycle is presented in [8], in which, based on graph theory, a model of threats that arise when managing an information security system was developed.

The need to use different models is due not only to different approaches to building a secure system, but also to different aspects of information security due to the difference in target objects [9].

Thus, the purpose of this work is to create a model of the life cycle of a secure chip during development and during operation by the end user, as well as providing a list of threats based on these models, depending on the security goals - confidentiality and integrity.

The models considered in the work do not include security threats arising in the framework of personnel management (their training or education) or document management (in physical or electronic form).

**Building a life cycle model**

The life cycle of a microcircuit is called all stages of the design and use of a microcircuit, from the stage of requirements formation to the stage of disposal [10-11].

Figure 1 shows a model of the life cycle of a microcircuit during its development. This model describes the development of hardware and firmware, since the development processes are similar.

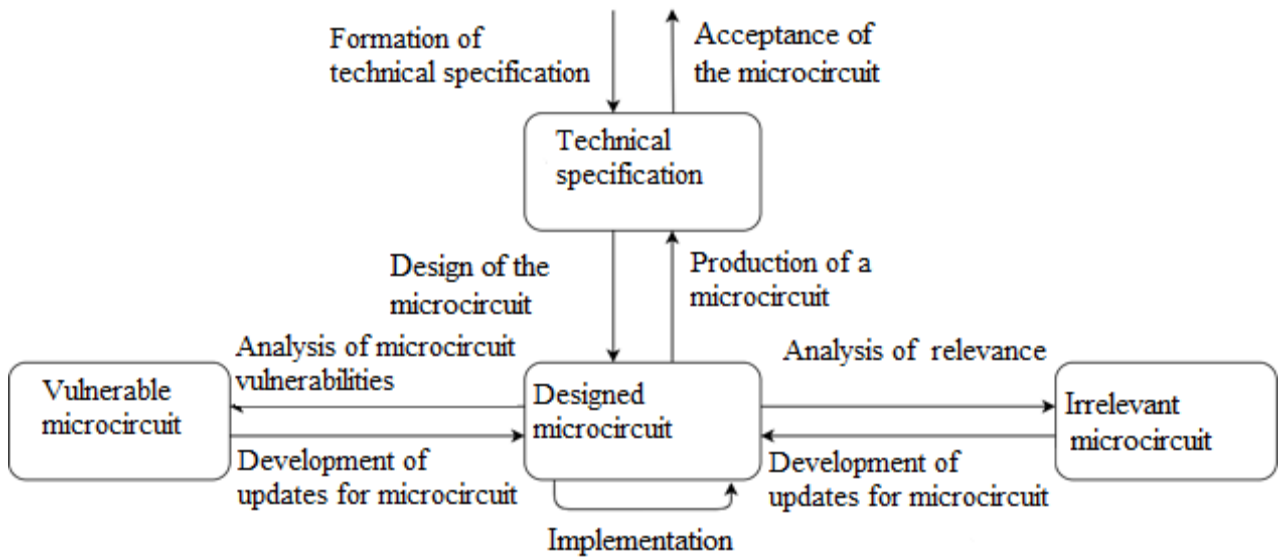

Rice. 1.Microcircuit development lifecycle

Stage 1.1 "Formation of terms of reference". The initial stage of the microcircuit development life cycle is the definition of formal requirements for device characteristics and software. At this stage, it is important to ensure the development of a threat model of the target device, including the definition of the intended purpose of the device.

Stage 1.2 "Design of the microcircuit". This stage includes a set of processes related to the development and debugging of functional blocks of the microcircuit. To create a microcircuit design, computer-aided design tools are used, and testing of the resulting blocks is carried out using modeling tools [12].

This stage is time-consuming due to the need for long-term planning, taking into account many potential problems and opportunities.

Stage 1.3 "Implementation". At this stage, PCB prototyping takes place , a software development kit for the microcircuit is created ( Software Development Kit , SDK), writing the source code for software that will run on the chip being developed. The software is also tested and debugged by simulating the developed microcircuit on the FPGA.

Stage 1.4 "Analysis of the microcircuit for the presence of vulnerabilities." At this stage, the functional blocks of the microcircuit are checked and the software is analyzed for compliance with safety requirements.

Stage 1.5 "Development of software and hardware patches for the microcircuit". At the stage of development of microcircuit patches , software versions and functional blocks of the microcircuit are adjusted taking into account the identified vulnerabilities in order to avoid the risks of information security threats.

Stage 1.6 "Analysis of the relevance of the microcircuit". At the stage of analyzing the relevance of the microcircuit, the support of modern protocols, operating systems and protection measures in the functional blocks of the microcircuit and in the software is checked.

Stage 1.7 "Development of software and hardware updates for the microcircuit". At this stage, the software and hardware of the microcircuit are updated by adding support for modern protocols, operating systems and protection measures to the microcircuit and software.

Stage 1.8 "Production of a microcircuit". This stage is a set of processes that must be performed to obtain a finished microcircuit. These processes include the steps involved in converting semiconductor materials into silicon wafers, fabricating masks containing topology images that will be transferred to silicon wafers after irradiation with ultraviolet light to form an integrated circuit (IC), separating the die from the silicon wafer, and packing the die into a physical container. . At this stage, it is also possible to initialize program modules that are not directly related to the logic of user applications (for example, the loader).

Stage 1.9 "Acceptance of the microcircuit". This stage is the final stage of the development of the chip and involves the performance of a full comprehensive testing of both functional blocks in particular and the chip as a whole to verify and confirm that the implemented chip and related software (SDK) meet the requirements for functionality and performance.

Figure 2 shows a model of the life cycle of a microcircuit during its operation by the end user. This model is suitable for describing both developers-integrators of a microcircuit in a new device, and end users, for the reason that the selected operation processes are similar.

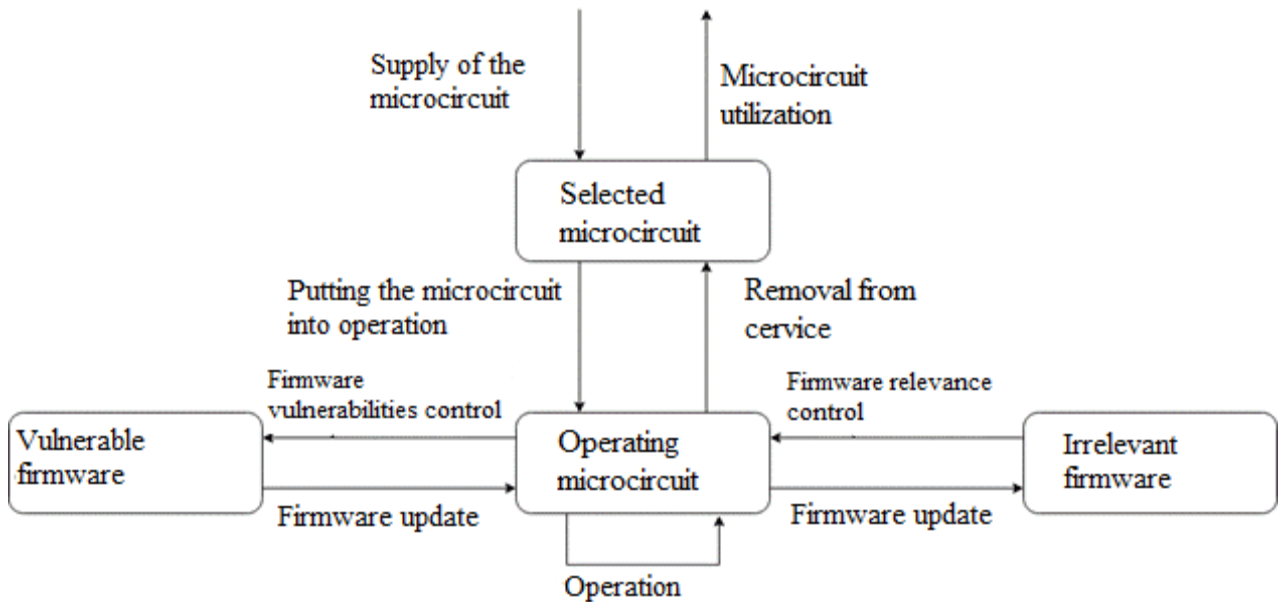

Rice. 2. The life cycle of the operation of the microcircuit by the end user

Stage 2.1 "Supply of the microcircuit". Based on the requirements of the target device, a selection of devices is made based on performance and security criteria.

Stage 2.2 "Putting the microcircuit into operation". This stage involves the execution of procedures for safely installing the selected microcircuit into the operating environment (for example, CII devices), installing firmware and bringing the device into a functioning state.

Stage 2.3 "Operation". At the operational stage, the microcircuit is used at the discretion of the user, which can be either the developer of a new device or its end user.

Stage 2.4 "Firmware control for the absence of vulnerabilities". The purpose of the presented stage is to search, identify and analyze vulnerabilities inherent in the current version of the firmware of the microcircuit and eliminated by applying a corrective patch .

Step 2.5 "Installing firmware patches ". At this stage, changes that were released by the developer of the chip or end device are applied to the firmware of the microcircuit [13]. The changes correct existing bugs that contribute to the emergence of threats in the chip or device.

Stage 2.6 "Checking the relevance of the firmware". At this stage, an analysis is made of the degree of obsolescence of the firmware of the chip in order to identify components that need to be updated.

Step 2.7 Install Firmware Updates. The current phase is to update the chip's firmware content to add new functionality to meet current security standards, protocols, and operating systems.

Stage 2.8 "Removal of the microcircuit from service". At this stage, all stored encryption keys [14], user information, and device configuration data are deleted. Deletion occurs by completely erasing the contents of the memory and resetting the device to factory settings so that it can be reused if necessary.

Stage 2.9 "Utilization of the microcircuit". At this stage, the chip or device with the chip in the composition is physically destroyed.

Table 1 and Table 2 present the threats, confidentiality and integrity, respectively, inherent in the stages during the development and operation of the chip. Some of the presented threats correspond to the threats from the list "GOST R 58412-2019" [15].

Table 1
**Microchip privacy threat**

| Stage number | Development | Exploitation |
|---|---|---|
| 1 | The threat of identifying microcircuit vulnerabilities due to the disclosure of information about the security requirements for the created microcircuit (corresponds to 5.1.2 of GOST R 58412-2019) | The threat of revealing microcircuit vulnerabilities microcircuit in the process of its delivery |
| 2 | The threat of revealing microcircuit vulnerabilities due to the disclosure of information about the design of the microcircuit architecture (corresponds to 5.2.2 of GOST R 58412-2019) | The threat of revealing microcircuit vulnerabilities due to the disclosure of information about security parameters, including encryption keys |
| 3 | The threat of revealing vulnerabilities of the microcircuit due to the disclosure of the source code of the firmware of the microcircuit or the circuit of the microcircuit (corresponds to 5.3.5 of GOST R 58412-2019) | The threat of revealing microcircuit vulnerabilities due to the disclosure of information about violation of the rules for the operation of the microcircuit |
| four | The threat of revealing microcircuit vulnerabilities due to the disclosure of information about testing the microcircuit and its firmware for vulnerability (corresponds to 5.4.2 of GOST R 58412-2019) | The threat of identifying microcircuit vulnerabilities due to the lack of control over updates for vulnerabilities |
| five | The threat of revealing vulnerabilities of the microcircuit due to the disclosure of information on the development of software and hardware patches of the microcircuit | Vulnerability exposure threat of firmware updates |
| 6 | The threat of identifying microcircuit vulnerabilities due to the disclosure of information about obsolete protocols eliminated by the patch , OS and protection measures in the functional blocks of the microcircuit and in firmware | The threat of revealing microcircuit vulnerabilities due to the lack of control over the presence of updates to the firmware |
| 7 | The threat of revealing vulnerabilities of the microcircuit due to the disclosure of information about the development of software and hardware updates of the microcircuit | Threat of detection of microcircuit vulnerabilities due to disclosure of information about violation of the rules for installing microchip firmware updates |
| 8 | The threat of revealing microcircuit vulnerabilities due to the disclosure of information about the technical process of the microcircuit | The threat of revealing microcircuit vulnerabilities due to the disclosure of information about the configuration of the device when decommissioning it |
| nine | The threat of revealing microcircuit vulnerabilities due to the disclosure of information about software errors and program vulnerabilities (corresponds to 5.6.2 of GOST R 58412-2019) | The threat of revealing microcircuit vulnerabilities due to the disclosure of information about the circuitry of the device due to its incomplete utilization |

table 2
**Threat to the integrity of the microcircuit**

| Stage number | Development | Exploitation |
|---|---|---|
| 1 | The threat of microcircuit vulnerabilities due to errors made when setting the security requirements for the microcircuit being developed (corresponds to 5.1.1 from GOST R 58412-2019) | The threat of introducing vulnerabilities into the microcircuit during its delivery (corresponds to 5.5.1 of GOST R 58412-2019) |
| 2 | The threat of microcircuit vulnerabilities due to errors made when creating the design of the architecture of the microcircuit functional blocks and in the firmware (corresponds to 5.2.1 from GOST R 58412-2019) | Threat of microcircuit firmware vulnerabilities due to errors made during firmware installation (corresponds to 5.2.1 of GOST R 58412-2019) |
| 3 | The threat of introducing vulnerabilities into the source code of the firmware and into the functional blocks of the microcircuit during its development (corresponds to 5.3.1 of GOST R 58412-2019) | The threat of introducing vulnerabilities into the program when managing the software configuration (corresponds to 5.7.1 of GOST R 58412-2019) |
| four | The threat of the appearance of program vulnerabilities due to errors when performing software testing (corresponds to 5.4.3 of GOST R 58412-2019) | Threat of exploiting microcircuit vulnerabilities due to violations of the rules for monitoring firmware for the absence of vulnerabilities |
| five | Threat of non-correction of detected program vulnerabilities (corresponds to 5.6.1 of GOST R 58412-2019 ) | The threat of introducing vulnerabilities into a software patch (corresponds to 5.5.3 of GOST R 58412-2019) |
| 6 | The threat of vulnerabilities due to the lack of control over the relevance of third-party obsolete software components used | The threat of exploiting microcircuit vulnerabilities due to violation of the rules for monitoring the relevance of firmware |
| 7 | The threat of introducing program vulnerabilities through the use of vulnerable components borrowed from third-party software developers (corresponds to 5.3.2 of GOST R 58412-2019) | Threat of microchip vulnerabilities due to violation of firmware update rules |
| 8 | The threat of introducing program vulnerabilities due to incorrect use of tools in software development (corresponds to 5.3.3 of GOST R 58412-2019) | Threat of chip vulnerabilities due to lack of support or obsolescence of the chip (refusal to decommission) |
| nine | The threat of introducing vulnerabilities into the firmware source code during its acceptance (corresponds to 5.3.1 of GOST R 58412-2019) | Threat of reuse of chip components due to incomplete recycling |

The relevance of the above list of threats is justified by the fact that it expands the list of threats presented in GOST R 58412-2019, in addition to 19 threats from the standard, more than 15 threats are proposed in the work. This is due to the approach used in the article, which consists in dividing the life cycle processes into the development and operation of the microcircuit, as well as dividing the life cycle processes depending on security goals, which makes it possible to identify narrowly targeted threats.

Threats from the standard are too general and in most cases cannot be applied in real projects. For example, in GOST, the presented threat "Threat of introducing vulnerabilities into software updates" does not reflect the nature of the threat during development and during operation.

**Conclusion**

This article presented the life cycle of secure microcircuits, described its main stages, and also considered the threats inherent in each stage, depending on the security goals - threats to the confidentiality and integrity of the microcircuit and its firmware. For each stage (chip development and operation of the chip), 9 confidentiality and integrity threats are identified that apply to any chip (36 threats in total).

The proposed approach to the formation of a list of threats, based on typical stages of the life cycle of a security system, has a number of advantages. In particular, it not only formalizes the list of threats presented in GOST R 58412-2019, but also supplements it.

This research was funded by the Ministry of Science and Higher Education of Russia, Government Order for 2020–2022, Project No . FEWM-2020-0037 (TUSUR).

*Literature*

1. Bhaiyat H. The Emergence of IIoT and its Cyber Security Issues in Critical Information Infrastructure / H. Bhaiyat , S. Sithungu // eccws . - 2022. - Vol. 21, No 1. - P. 46-51.

2. Noseda M. Performance Analysis of Secure Elements for IoT / M. Noseda ; L. Zimmerli ; T. Schläpfer ; A. Rust // IoT . - 2021. - Vol. 3, No 1.–P. 1–28.

3. Ebad SA Exploring How to Apply Secure Software Design Principles // IEEE Access. - 2022. - №10. - S. _ 128983 -128993.

4. Intent-Driven Secure System Design: Methodology and Implementation / SE Ooi , R. Beuran , T. Kuroda, T. Kuwahara , R. Hotchi , N. Fujita, Y. Tan // Computers & Security - 2022 - Vol. 124 - P. 102955.

5. Threat Model for IoT Systems on the Example of OpenUNB Protocol / A. Shelupanov , A. Konev, T. Kosachenko, D. Dudkin // International Journal of Emerging Trends in Engineering Research. - 2019. - Vol. 7, No 9. -P . 283-290.

6. AA Konev, TE Mineeva , ML Soloviev , AA Shelupanov , and MP Silich , "Model of the life cycle of the information security system," Bezopasnost informacionnyh technology , 2018, vol. 25, no. 4. National Research Nuclear University MEPhI (Moscow Engineering Physics Institute), pp. 34–42. (In Russian).

7. Alenezi M. Security Risks in the Software Development Lifecycle / M. Alenezi S. Almuairfi // International Journal of Recent Technology and Engineering (IJRTE). Blue Eyes Intelligence Engineering and Sciences Engineering and Sciences Publication (BEIESP). - 2019. - Vol. 8, No 3. - P. 7048-7055.

8. ML Soloviev et al., "Model of security threats arising from the management of information security systems," Proceedings of Tomsk State University of Control Systems and Radioelectronics , 2019, vol. 22, no. 3. Tomsk State University of Control Systems and Radioelectronics (TUSUR), pp. 31–36. (In Russian).

9. Computer network threat modeling / A. Novokhrestov , A. Konev, A. Shelupanov , A. Buymov // IOP Conf. Series: Journal of Physics: Conf. Series. - 2020. - Vol. 1488, No 1. - P. 6 .

10. Yousefnezhad N. Security in product lifecycle of IoT devices: A survey / N. Yousefnezhad , A. Malhi , K. Främling // Journal of Network and Computer Applications. - 2020. - Vol. 171. - P. 102779.

11. GOST R 57193-2016. System lifecycle processes. Available at: https://docs.cntd.ru/document/1200141163 (Accessed: November 10, 2022). (In Russian).

12. Chan C.-H. Trending IC design directions in 2022 / C.-H. Chan, L. Cheng, W. Deng, P. Feng, L. Geng , M. Huang, H. Jia , L. Jie , K.-M. Lei, X. Liu, X. Liu, Y. Liu, Y. Lu, K. Nie , D. Pan, N. Qi, S.-W. Sin, N. Sun // Journal of Semiconductors. IOP Publishing. - 2022. - Vol. 43, No 7. - P. 071401.

13. El Jaouhari S. Secure firmware Over-The-Air updates for IoT / S. El Jaouhari , E. Bouvet / Survey, challenges, and discussions // Internet of Things. - 2022. - Vol. 18. - P. 100508.

14. Mathur S. Internet of Things ( IoT ) and PKI-Based Security Architecture / S. Mathur , A. Arora // Industrial Internet of Things and Cyber-Physical Systems. – 2020. – P. 25–46.

15. GOST R 56939-2016. Threats to information security in software development Available at: https://docs.cntd.ru/document/12001355258 (Accessed: November 10, 2022). (In Russian).